\documentclass{article}

\usepackage{PRIMEarxiv}

\usepackage[utf8]{inputenc} 
\usepackage[T1]{fontenc}    
\usepackage{hyperref}       
\usepackage{url}            
\usepackage{booktabs}       
\usepackage{amsfonts}       
\usepackage{nicefrac}       
\usepackage{microtype}      
\usepackage{lipsum}
\usepackage{fancyhdr}       
\usepackage{graphicx}       
\graphicspath{{media/}}     
\usepackage{longtable}
\usepackage{xltabular}
\usepackage[nolist]{acronym}
\usepackage{multirow}
\usepackage{xcolor}
\usepackage[numbers]{natbib}

\newcommand{\origin}[1]{\newline {\small#1}}
\newcommand{\dataset}[2]{\textbf{#1} \origin{#2} \vspace{0.2cm}}
\newcommand{\myHref}[1]{\href{#1}{\underline{Link}}}
\newcommand{\myDomAppSig}[3]{\textbf{#1} \newline {#2} \newline Signals: {#3} \vspace{0.2cm}}
\newcommand{\myRow}[8]{\textbf{#1} & \dataset{#2}{#3} & \myDomAppSig{#5}{#6}{#8} & {#7} & \myHref{#4} \\}
\newcommand{\sumDataSets}{{110}}
\newcommand{\nrDataSetsInClassifiedDomains}{{11}}
\newcommand{\nrDataSetsInClassifiedApplications}{{43}}
\newcommand{\nrDataSetsFaultDetection}{{18}}
\newcommand{\nrDataSetsDiagnosis}{{35}}
\newcommand{\nrDataSetsHealthAssessment}{{3}}
\newcommand{\nrDataSetsPrognosis}{{54}}
\newcommand{\sumSignals}{{22}}
\newcommand{\ie}{i.e.,~}
\newcommand{\eg}{e.g.,~}

\acrodef{AI}[AI]{Artificial Intelligence}

\acrodef{1DCNN}[1D-CNN]{one-dimensional convolutional neural network}
\acrodef{D}[D]{diagnosis}
\acrodef{ES}[ES]{engineering system}

\acrodef{FD}[FD]{fault detection}

\acrodef{HA}[HA]{health assessment}
\acrodef{ML}[ML]{Machine Learning}

\acrodef{NIST}[NIST]{National Institute of Standards and Technology}
\acrodef{NASA}[NASA]{National Aeronautics and Space Administration}
\acrodef{PHM}[PHM]{Prognostics and Health Management}
\acrodef{PIML}[PIML]{Physics-informed Machine Learning}
\acrodef{P}[P]{prognosis}
\acrodef{RUL}[RUL]{Remaining Useful Life}
\acrodef{TRF}[TRF]{Trust Region Reflective}
\acrodef{TGDS}[TGDS]{Theory-guided Data Science}
\acrodef{TSFEL}[TSFEL]{Time Series Feature Extraction Library}
\acrodef{UCI}[UCI]{University of California Irvine}

\title{Overview of publicly available degradation data sets for tasks within prognostics and health management
\thanks{\textit{\underline{Citation \cite{Mauthe.2025}}}: 
{Fabian Mauthe, Luca Steinmann, Moritz Neu, and Peter Zeiler. Overview and analysis of publicly available degradation data sets for tasks within prognostics and health management. In \textit{Proceedings of the 35th European Safety and Reliability Conference and the 33rd Society for Risk Analysis Europe Conference}, pages 945–952, Singapore, 2025. Research Publishing.}} 
\thanks{\textit{\underline{Citation \cite{hagmeyer2021creation}}}: 
{Simon Hagmeyer, Fabian Mauthe, and Peter Zeiler. Creation of publicly available data sets for prognostics and diagnostics addressing data scenarios relevant to industrial applications. \textit{International Journal of Prognostics and Health Management}, 12(2):1–20, 2021.}} 
}

\author{
  Fabian Mauthe \\
  Institute for Technical Reliability and Prognostics IZP, Esslingen University of Applied Sciences\\
  Goeppingen, Germany \\
  \texttt{fabian.mauthe@hs-esslingen.de} \\
   \And
    Christopher Braun \\
  Institute of Industrial Manufacturing and Management IFF, University of Stuttgart \\
  Stuttgart, Germany\\
  \texttt{christopher.braun@iff.uni-stuttgart.de} \\
  \And
  Julian Raible \\
  Institute of Industrial Manufacturing and Management IFF, University of Stuttgart \\
  Stuttgart, Germany\\
  \texttt{julian.raible@iff.uni-stuttgart.de} \\
   \And
  Peter Zeiler \\
  Institute for Technical Reliability and Prognostics IZP, Esslingen University of Applied Sciences\\
  Goeppingen, Germany \\
  \texttt{peter.zeiler@hs-esslingen.de} \\
  \And
  Marco F. Huber \\
  Institute of Industrial Manufacturing and Management IFF, University of Stuttgart \\
  Fraunhofer Institute of Manufacturing Engineering and Automation IPA \\  
  Stuttgart, Germany\\
  \texttt{marco.huber@ieee.org} \\
}

\begin{document}
\maketitle



\section{Prognostics and Health Management}
In the realm of modern manufacturing, significant emphasis is placed on improving the reliability, performance, and service life of complex \acp{ES}, which has catalyzed the emergence of \ac{PHM} as an essential discipline in the past two decades \cite{Zio.2022,Lei.2018}. \ac{PHM} includes a range of methods dedicated to the early detection, prognosis, and mitigation of impending faults or failures in \acp{ES}, thereby facilitating sophisticated maintenance strategies and enhancing operational efficiency \cite{Atamuradov.2017}.

Central to the efficacy of \ac{PHM} methods is the acquisition and analysis of degradation data, which encapsulates the evolving health condition of \acp{ES} over time \cite{Atamuradov.2017}. Degradation data serves as a rich source of information, offering invaluable insights into the underlying degradation processes, failure modes, and performance trends of \acp{ES} \cite{Lei.2018,hagmeyer2021creation}. 
The availability of appropriate degradation data remains a significant challenge in both industrial practice and academic research \cite{Zio.2022}. Consequently, publicly available degradation data sets are of considerable value. They facilitate the development and empirical benchmarking of \ac{PHM} methodologies and serve as a basis for demonstrating the practical applicability of the proposed approaches \cite{Mauthe.2025FOR,Zoller.2023,Ramasso.2014}.

Although publicly available degradation data sets are of significant importance, the literature addresses this subject to a limited extent. Overviews often consider a restricted range of sources and platforms or only specific applications, resulting in constrained summaries \cite{Soualhi.2023paper,Su.2024}. Analyses usually emphasize the amount of data or data quality, overlooking specific \ac{PHM} aspects \cite{Jourdan.2021}. As a result, the search process for suitable data sets is often very time-consuming for users of \ac{PHM} methods \cite{Mauthe.2025}. 
Therefore, the objective of this work is to provide a comprehensive overview of publicly available degradation data sets and a detailed \ac{PHM}-specific analysis. To do this, the analysis presented in \cite{Mauthe.2025} is used. This \ac{PHM}-specific analysis covers aspects that are pertinent to both users of \ac{PHM} methods and of corresponding data sets. These aspects comprise the application represented in the data set, the originating domain of this application, the specific task the data set is designed to address, and the types of signals contained in the data set \cite{Mauthe.2025}.

In the remainder of this work, the tasks of diagnostics and prognostics are listed in Section~\ref{section:tasks}, and an overview of publicly available degradation data sets along with a \ac{PHM}-specific analysis is presented in Section~\ref{section:data_sets}. Section~\ref{sec:conclusion} concludes the results of this work. The authors aim to update the overview as well as the analysis regularly in the future, as new data sets are continually being published in the dynamic research field of \ac{PHM}.

\section{Tasks of Diagnostics and Prognostics}
\label{section:tasks}
Extensive and representative degradation data address a specific task of diagnostics and prognostics within \ac{PHM}. Based on~\cite{Zio.2022}, \cite{Su.2024}, and \cite{hagmeyer2021creation}, these tasks can be subdivided as follows:
\begin{itemize}
    \item \textbf{Fault detection/anomaly detection}: Detect a fault state or anomaly in an \ac{ES} without considering its root cause. This results in a binary classification problem with the states fault or no fault.
    \item \textbf{Diagnosis}: Assign one or more causes to a detected fault state.
    \item \textbf{Health assessment}: Assess the state of degradation or the current risk of failure of an \ac{ES} based on its current condition.
    \item \textbf{Prognosis}: Prediction of future degradation behavior or the current \ac{RUL}.
\end{itemize}

This list of tasks can also be interpreted as a sequential framework for implementing a comprehensive \ac{PHM} application, starting with fault detection and ending with prognosis \cite{Zio.2022}.
This can be illustrated by the following example of rolling bearing failure: Fault detection identifies an abnormal vibration pattern. Diagnosis then determines the root cause, for instance, a specific bearing component. The health assessment quantifies how this fault affects the bearing’s condition, while the prognosis predicts the bearing’s \ac{RUL}. \cite{Mauthe.2025}


\section{Data Sets for Diagnostics and Prognostics}
\label{section:data_sets}
\subsection{Overview}
Due to regular data challenges and several research activities of the \ac{PHM} community in recent years, new data sets are continuously being published. Therefore, the overview by~\citet{Mauthe.2025} (which is already based on the overview by \citet{hagmeyer2021creation}) is updated using a similar procedure, \ie the identical scope of platforms and sources of degradation data sets is considered \cite{hagmeyer2021creation}. As a result, the overview is extended to \sumDataSets~data sets in total.
In accordance with \cite{hagmeyer2021creation} and \cite{Mauthe.2025}, data sets that focus solely on process quality without including the degradation state are excluded from consideration.
Table~\ref{tab:list_data_sets} in \emph{\nameref{appendix_overview_data_sets}} contains the extended overview of publicly available data sets for the tasks of diagnostics and prognostics within \ac{PHM}, sorted alphabetically.

\subsection{Data Set Analysis}
The detailed \ac{PHM}-specific analysis of the data sets is based on the procedure presented in \cite{Mauthe.2025}. The analysis is intended to shorten the time-consuming search for suitable data sets for its users. In particular, the focus of this analysis is on the following \ac{PHM}-specific aspects \cite{Mauthe.2025}:
\begin{itemize}
    \item \textbf{Application}: The application within a data set refers to the object of consideration from which the data originate. Thus, an application can represent a single component, such as a bearing, or an entire system, such as a production line.
    \item \textbf{Domain}: The domain from which the application originates. The domains associated with the previous examples would then be, for example, mechanical components or production systems.
    \item \textbf{Task}: The task, which is addressed by the corresponding data set. A distinction is made among the tasks described in Section~\ref{section:tasks}: fault detection, diagnosis, health assessment, and prognosis.
    \item \textbf{Signal}: The types of signals included in the data set. This refers to signals that offer insight into the health state of the specific application or can be derived from it and used within PHM methods.
\end{itemize}
Based on these PHM-specific aspects, \citet{Mauthe.2025} developed a taxonomy that is tailored to the classification of data sets in a PHM-specific manner. This taxonomy categorizes all data sets based on their application and domain of origin, the task addressed by the data set, and the relevant signals contained in the data set.
The results using this taxonomy for the \sumDataSets~publicly available data sets are provided in Table~\ref{table:taxonomyPubliclyDataSets} in \emph{\nameref{appendix_taxonomy}} and in Table~\ref{table:taxonomySignalPubliclyDataSets} in \emph{\nameref{appendix_signals_data_sets}}. The respective assignment of the \ac{PHM}-specific aspects (application, domain, task, and signal) to the individual data sets can also be found in Table~\ref{tab:list_data_sets} in \emph{\nameref{appendix_overview_data_sets}}.

\paragraph{Domains and applications:}
\label{subsection:applications_and_domains}
Applying the presented taxonomy, the \sumDataSets~publicly available degradation data sets are classified into \nrDataSetsInClassifiedDomains~domains (see Table~\ref{table:taxonomyPubliclyDataSets}). The mechanical component domain contains the most data sets ({26}), followed by the electrical component domain ({25}) and the drive technology domain ({12}). Apart from these three largest domains, the remaining domains contain between {two and eight} data sets. Moreover, {five} data sets have unknown applications. As the respective data sets can still be used to apply \ac{PHM} methods, they are taken into account and assigned to the unknown domain.

In total, \nrDataSetsInClassifiedApplications~different applications are included in the \nrDataSetsInClassifiedDomains~domains, covering a variety of degradation processes. However, as these applications are spread across \sumDataSets~data sets, the number of available data sets per application is limited. This is reflected in a more in-depth consideration of Table~\ref{table:taxonomyPubliclyDataSets}: The applications bearing ({17} data sets) and battery ({16} data sets) are considered most frequently, originating from the two main domains, mechanical component and electrical component, respectively. The next most frequent applications are aircraft engine, filtration, and production line, each occurring in {five} data sets. The limited number of data sets per application is further highlighted by the mechatronic system domain, which contains the most different applications ({eight}), with only one data set per application. Overall, only {nine} applications are covered by three or more data sets.

\paragraph{Considered tasks within data sets:}
\label{subsection:tasksindatasets}
The diagnostic and prognostic tasks of the data sets\footnote{Certain data sets may be assigned to several tasks. As these tasks build on each other, the highest-ranking task is assigned to a respective data set. Therefore, for each data set, only one task is assigned.}, as introduced in Section~\ref{section:tasks}, are shown in the third column of Table~\ref{table:taxonomyPubliclyDataSets}. With \nrDataSetsPrognosis~assigned data sets, the prognosis task is the most represented, followed by diagnosis with \nrDataSetsDiagnosis~entries, whereas fault detection is assigned to \nrDataSetsFaultDetection~data sets. As in the previous classification, the transition from diagnosis to health assessment, as well as the transition to prognosis, is often not distinct. Accordingly, only {three} data sets explicitly address the task of health assessment.

For prognosis, the electrical component domain contributes {22}~out of \nrDataSetsPrognosis~data sets, of which {15}~address battery applications. The mechanical component and drive technology domains follow with eight and seven prognosis data sets, respectively. Bearing is the second most common application for prognosis ({seven} data sets), while aircraft engine and filtration each occur in {five} prognosis data sets. In total, {20}~distinct applications are available for prognosis.

For diagnosis, the mechanical component domain contains the most data sets ({13} out of \nrDataSetsDiagnosis), while the robotic and mechatronic system domains contain {five} diagnosis data sets each, and the production system domain contains {three}. The bearing application is the most frequently used for diagnosis ({eight} data sets), followed by articulated robot applications ({four}) and both gear and production line applications ({three} each). In general, {21}~different applications are considered in the \nrDataSetsDiagnosis~data sets for diagnosis, indicating a wide variety of applications.

For fault detection, the mechanical component and production system domains contribute the largest shares, with {five} and {four} data sets, respectively. Drive technology and manufacturing process contribute {two} fault detection data sets each, and the unknown domain contains {two} fault detection data sets without a specified application. Regarding applications, only bearing, milling, and production line occur in {two} data sets each; all remaining applications occur only once.

\paragraph{Signals within data sets:}
\label{subsection:signalsindatasets}
Table~\ref{table:taxonomySignalPubliclyDataSets} summarizes the \sumSignals~signals used in at least two of the given data sets. Note that Table~\ref{table:taxonomySignalPubliclyDataSets} only lists signals that are measured directly. Values calculated from these, such as capacity or power, are excluded. Nevertheless, if only calculated signals are included in a data set, they are still listed. Also summarized are signals of a comparable nature, such as speed, including rotation and velocity. 

Vibration appears most frequently, occurring in {34}~data sets, followed by current and temperature with {33}~data sets each, and voltage with {25}~data sets. These signals are typically measured within electrical and mechanical component domains (\eg battery and bearing applications) and therefore are most prevalent in the analyzed data sets. Speed and pressure are also common, appearing in {20}~and {18}~data sets, respectively. Anonymized signals and flow rate occur in {ten} data sets each; however, anonymized signals permit specific observations mainly within a data set or application, limiting their utility for general analysis. Regarding the availability of data sets per signal, a limitation similar to applications can be observed: {ten} of the \sumSignals~signals occur in {four or fewer} data sets.

Across tasks, signal occurrences show characteristic differences. For prognosis, current ({22}) and temperature ({21}) are most prevalent, followed by voltage ({19}) and vibration ({14}). For diagnosis, vibration dominates ({16}), followed by speed ({11}), current ({nine}), and temperature ({eight}), indicating that these signals are also relevant beyond purely electrical component data sets.

\section{Conclusion}
\label{sec:conclusion}
This work updates an overview of publicly available degradation data sets for tasks of diagnostics and prognostics within \ac{PHM} and structures them using a dedicated taxonomy. The classification makes the current coverage across domains, applications, and tasks explicit and highlights where publicly available data are concentrated versus where only sparse coverage exists. Overall, the results underline that benchmarking is well supported for a few frequently studied use cases and signal modalities, while broader generalization and task-specific evaluations remain limited by uneven coverage. Maintaining and extending this overview will help track how coverage evolves and where additional publicly available data would have the greatest impact.

\bibliographystyle{unsrtnat}  
\bibliography{references}

\section*{Appendix A: Overview of Publicly Available Data Sets}
\label{appendix_overview_data_sets}

The designation of the data sets is divided into two parts, consisting of the source or platform and the title under which they were published. In addition, the overview contains the following columns:
\begin{itemize}
    \item Domain, Application, and Signals: The associated domain from which the data set originates and the respective application within this domain. In addition, the signals present in the respective data set are listed.
    \item Task: The corresponding task within \ac{PHM} addressable with this data set.
    \item URL: The respective hyperlink to the source or platform hosting the data set.
\end{itemize}

\begin{xltabular}{\textwidth}{|p{0.6cm}|X|p{4.1cm}|p{2.3cm}|p{1.0cm}|}
    \caption{Overview of publicly available data sets}
    \label{tab:list_data_sets} \\
    
    \toprule
	\textbf{No.} & {\textbf{Data Set Designation} \newline Reference to Data Origin} & {\textbf{Domain} \newline Application and Signal} & \textbf{Task} & \textbf{URL}\\ 
    \midrule
    \endfirsthead

    \multicolumn{5}{c}{\textit{\tablename\ \thetable{} -- continued from previous page}} \\
    \toprule
    \textbf{No.} & {\textbf{Data Set Designation} \newline Reference to Data Origin} & {\textbf{Domain} \newline Application and Signal} & \textbf{Task} & \textbf{URL}\\ 
    \midrule
    \endhead
    
    \midrule \multicolumn{5}{|r|}{{\textit{Continued on next page}}} \\ \hline
    \endfoot

    \bottomrule
    \endlastfoot

\myRow{1}{4TU - Lifecycle ageing tests on commercial 18650 Li ion cell
}{Trad, Khiem (2021): Lifecycle ageing tests on commercial 18650 Li ion cell @ 25°C and 45°C. Version 1. \textit{4TU.ResearchData}. dataset. https://doi.org/10.4121/13739296.v1}{https://data.4tu.nl/articles/_/13739296/1}{Electrical component}{Battery}{Prognosis}{current, temperature, voltage}

\myRow{2}{4TU - Motor Current and Vibration Monitoring Dataset}{Bruinsma, Sietze; Geertsma, Rinze; Loendersloot, Richard; Tinga, Tiedo (2024): Motor Current and Vibration Monitoring Dataset for various Faults in an E-motor-driven Centrifugal Pump. doi: 10.4121/2b61183ec14f-4131-829b-cc4822c369d0.v3}{https://data.4tu.nl/datasets/2b61183e-c14f-4131-829b-cc4822c369d0}{Process technology}{Centrifugal pumps}{Diagnosis}{current, vibration, voltage}

\myRow{3}{AIDAR Lab - Air Compressor}{Nishchal K. Verma, R. K. Sevakula,
S. Dixit and Salour A. (2016). Intelligent Condition Based Monitoring using Acoustic Signals for Air Compressors, \textit{IEEE Transactions on Reliability}, vol. 65, no. 1, pp. 291-309.}{http://www.iitk.ac.in/idea/datasets/}{Mechatronic system}{Air Compressor}{Diagnosis}{acoustic emission}

\myRow{4}{AIDAR Lab - Drill Bit}{Nishchal K. Verma, R. K. Sevakula, S. Dixit and Salour A. (2015). Data Driven Approach for Drill Bit Monitoring, \textit{Reliability Digest}, pp. 19-26.}{http://www.iitk.ac.in/idea/datasets/}{Manufacturing process}{Drill bit}{Diagnosis}{vibration}

\myRow{5}{Aramis - Data Challenge ESREL2020PSAM15}{Francesco Cannarile (Aramis Srl, Italy), Michele Compare (Aramis Srl, Italy and Politecnico di Milano, Italy), Piero Baraldi (Politecnico di Milano, Italy), Zhe Yang (Politecnico di Milano, Italy), Enrico Zio (Politecnico di Milano, Italy and MINES ParisTech, France).}{https://aramis3d.com/innovation-challenges/}{Unknown}{Simulation (time-continuous stochastic process)*}{Fault detection}{vibration}

\myRow{6}{Backblaze - Hard Drive Stats}{Backblaze, Inc.}{https://www.backblaze.com/b2/hard-drive-test-data.html}{Drive technology}{Disk Drives}{Fault detection}{temperature}

\myRow{7}{Bearing Data Center - Fan and Bearing}{Case Western Reserve University, USA}{https://engineering.case.edu/bearingdatacenter/download-data-file}{Mechanical component}{Bearing}{Diagnosis}{acoustic emission}

\myRow{8}{Calce - Battery Data Repository}{CALCE Center for Advanced Life Cycle Engineering, University of Maryland, website: https://calce.umd.edu/battery-data}{https://calce.umd.edu/battery-data}{Electrical component}{Battery}{Prognosis}{current, temperature, voltage}

\myRow{9}{ETH Zurich - Run-to-Failure High-voltage Circuit Breaker Mechanical Test Dataset}{Hsu, C.-C. (2024). Run-to-failure high-voltage circuit breaker mechanical test dataset. \textit{ETH Z\"{u}rich Research Collection}. DOI: 10.3929/ethz-b-000676480}{https://doi.org/10.3929/ethz-b-000676480}{Electrical component}{Circuit breaker (high-voltage)}{Prognosis}{current, vibration}

\myRow{10}{GitHub - Motor current milling machine}{Apurv Rajeshkumar Darji (Developer), Dr. Mustafa Demetg\"{u}l (Academic Supervisor)}{https://github.com/ApurvDarji/thesis_wbk}{Manufacturing process}{Milling machine (motor current)}{Fault detection}{current}

\myRow{11}{GitHub - Predictive Maintenance using PySpark}{GitHub User: linya9191}{https://github.com/Azure/PySpark-Predictive-Maintenance}{Simulation}{Simulation*}{Fault detection}{unknown}

\myRow{12}{GitHub - XJTU-SY Bearing Datasets}{Biao Wang, Yaguo Lei, Naipeng Li, Ningbo Li (2020). A Hybrid Prognostics Approach for Estimating Remaining Useful Life of Rolling Element Bearings, \textit{IEEE Transactions on Reliability}, vol. 69, no. 1, pp. 401-412. DOI: 10.1109/TR.2018.2882682.}{https://github.com/WangBiaoXJTU/xjtu-sy-bearing-datasets}{Mechanical component}{Bearing}{Prognosis}{vibration}

\myRow{13}{Harvard Dataverse - Wind Turbine Main Bearing Fatigue Life Prediction}{Yucesan, Y. (2019). Wind Turbine Main Bearing Fatigue Life Prediction with PINN [Data set]. \textit{Harvard Dataverse}. https://doi.org/10.7910/DVN/ENNXLZ}{https://dataverse.harvard.edu/dataset.xhtml?persistentId=doi:10.7910/DVN/ENNXLZ}{Mechanical component}{Bearing}{Prognosis}{speed, temperature}

\myRow{14}{Kaggle - Air pressure system failures in Scania trucks}{Tony Lindgren and Jonas Biteus, Scania CV AB - Stockholm}{https://www.kaggle.com/uciml/aps-failure-at-scania-trucks-data-set}{Mechatronic system}{Air Pressure System}{Fault detection}{anonymized}

\myRow{15}{Kaggle - Bearings with Varying Degradation Behaviors}{Mauthe, F.; Hagmeyer, S.; Zeiler, P. (2025). Holistic simulation model of the temporal degradation of rolling bearings. In E. B. Abrahamsen, T. Aven, F. Bouder, R. Flage, and M. Yl\"{o}nen (Eds.), \textit{Proceedings of the 35th European Safety and Reliability Conference and the 33rd Society for Risk Analysis Europe Conference}. Research Publishing. DOI: 10.3850/978-981-94-3281-3\_ESREL-SRA-E2025-P8028-cd}{https://www.kaggle.com/datasets/prognosticshse/bearings-with-varying-degradation-behaviors}{Mechanical component}{Bearing (simulated)}{Prognosis}{vibration}

\myRow{16}{Kaggle - CNC Mill Tool Wear}{System-level Manufacturing and Automation Research Testbed (SMART) at the University of Michigan}{https://www.kaggle.com/datasets/shasun/tool-wear-detection-in-cnc-mill}{Manufacturing process}{Mill Tool}{Fault detection}{acceleration, current, feedrate, position, speed, voltage}

\myRow{17}{Kaggle - Condition Data with Random Recording Time}{Hagmeyer, S., Mauthe, F., Zeiler, P. (2021). Creation of Publicly Available Data Sets for Prognostics and Diagnostics Addressing Data Scenarios Relevant to Industrial Applications. \textit{International Journal of Prognostics and Health Management}, Volume 12, Issue 2, DOI: 10.36001/ijphm.2021.v12i2.3087}{https://www.kaggle.com/datasets/prognosticshse/condition-data-with-random-recording-time}{Process technology}{Filtration}{Prognosis}{flow rate, pressure}

\myRow{18}{Kaggle - E-coating ultrafiltration maintenance}{Tinsley Bridge Limited, Sheffield, United Kingdom}{https://www.kaggle.com/datasets/boyangs444/process-data-for-predictive-maintenance}{Manufacturing process}{Electrophoresis Painting}{Prognosis}{flow rate, pressure, temperature}

\myRow{19}{Kaggle - Genesis Demonstrator}{Institut f\"{u}r industrielle Informationstechnik (inIT) der Technischen Hochschule Ostwestfalen-Lippe (IMPROVE)}{https://www.kaggle.com/inIT-OWL/genesis-demonstrator-data-for-machine-learning}{Robotic}{Pick-and-Place Demonstrator}{Diagnosis}{acceleration, current, force, speed}

\myRow{20}{Kaggle - Microsoft Azure Predictive Maintenance}{Azure AI Notebooks for Predictive Maintenance}{https://www.kaggle.com/datasets/arnabbiswas1/microsoft-azure-predictive-maintenance}{Unknown}{Unknown Machine}{Prognosis}{pressure, speed, vibration, voltage}

\myRow{21}{Kaggle - One Year Industrial Component Degradation}{Institut f\"{u}r industrielle Informationstechnik (inIT) der Technischen Hochschule Ostwestfalen-Lippe (IMPROVE)}{https://www.kaggle.com/datasets/inIT-OWL/vega-shrinkwrapper-runtofailure-data?resource=download}{Production system}{Shrink-Wrapper}{Fault detection}{position, speed, torque}

\myRow{22}{Kaggle - Predictive Maintenance 1}{Creators Yuan Yao and Zhao Yuqi}{https://www.kaggle.com/c/predictive-maintenance1/overview}{Production system}{Log Data}{Fault detection}{unknown}

\myRow{23}{Kaggle - Preventive to Predictive Maintenance}{Hagmeyer, S., Mauthe, F., Zeiler, P. (2021). Creation of Publicly Available Data Sets for Prognostics and Diagnostics Addressing Data Scenarios Relevant to Industrial Applications. \textit{International Journal of Prognostics and Health Management}, Volume 12, Issue 2, DOI: 10.36001/ijphm.2021.v12i2.3087}{https://www.kaggle.com/prognosticshse/preventive-to-predicitve-maintenance}{Process technology}{Filtration}{Prognosis}{flow rate, pressure}

\myRow{24}{Kaggle - Production Plant Data for Condition Monitoring}{von Birgelen, Alexander; Buratti, Davide; Mager, Jens; Niggemann, Oliver (2018). Self-Organizing Maps for Anomaly Localization and Predictive Maintenance in Cyber-Physical Production Systems. \textit{Proceedings of the CIRP Conference on Manufacturing Systems (CIRP CMS 2018) CIRPCMS}}{https://www.kaggle.com/inIT-OWL/production-plant-data-for-condition-monitoring}{Production system}{Production line}{Fault detection}{anonymized}

\myRow{25}{Kaggle - Prognosis based on Varying Data Quality}{Hagmeyer, S., Mauthe, F., Zeiler, P. (2021). Creation of Publicly Available Data Sets for Prognostics and Diagnostics Addressing Data Scenarios Relevant to Industrial Applications. \textit{International Journal of Prognostics and Health Management}, Volume 12, Issue 2, DOI: 10.36001/ijphm.2021.v12i2.3087}{https://www.kaggle.com/datasets/prognosticshse/prognosis-based-on-varying-data-quality}{Process technology}{Filtration}{Prognosis}{dust feed, flow rate, pressure}

\myRow{26}{Kaggle - Pump}{Kaggle User: UnknownClass}{https://www.kaggle.com/datasets/anseldsouza/water-pump-rul-predictive-maintenance}{Process technology}{Water Pump}{Fault detection}{anonymized}

\myRow{27}{Kaggle - Sensor Fault Detection}{Schneider-Electric, France}{https://www.kaggle.com/datasets/arashnic/sensor-fault-detection-data}{Electrical component}{Temperature Sensor}{Fault detection}{temperature, voltage}

\myRow{28}{Kaggle - Similar System Data Set for Condition Prognosis}{Braig, M.; Zeiler, P. (2025): A Study on Using Transfer Learning to Utilize Information from Similar Systems for Data-Driven Condition Diagnosis and Prognosis. \textit{IEEE Access}, Volume 13, pp. 98485-98503, DOI: 10.1109/ACCESS.2025.3576435}{https://www.kaggle.com/datasets/prognosticshse/similar-system-data-set-for-condition-prognosis}{Process technology}{Filtration}{Prognosis}{flow rate, pressure}

\myRow{29}{Kaggle - Similar System Data Set for Fault Diagnosis}{Braig, M.; Zeiler, P. (2025): A Study on Using Transfer Learning to Utilize Information from Similar Systems for Data-Driven Condition Diagnosis and Prognosis. \textit{IEEE Access}, Volume 13, pp. 98485-98503, DOI: 10.1109/ACCESS.2025.3576435}{https://www.kaggle.com/datasets/prognosticshse/similar-system-data-set-for-fault-diagnosis}{Mechanical component}{Bearing}{Diagnosis}{vibration}

\myRow{30}{Kaggle - Versatile Production System}{Institut f\"{u}r industrielle Informationstechnik (inIT) der Technischen Hochschule Ostwestfalen-Lippe (IMPROVE)}{https://www.kaggle.com/inIT-OWL/versatileproductionsystem}{Production system}{Versatile Production System}{Fault detection}{control parameters}

\myRow{31}{Karlsruhe Institute of Technology - Ball Screw Drive Surface Defect Dataset}{Schlagenhauf, T. (2021): Ball Screw Drive Surface Defect Dataset for Classification. \textit{Karlsruhe Institute of Technology (KIT)}. DOI: 10.5445/IR/1000133819}{https://publikationen.bibliothek.kit.edu/1000133819}{Mechanical component}{Ball screw drive}{Fault detection}{image}

\myRow{32}{Karlsruhe Institute of Technology - Industrial Machine Tool Element Surface Defect Dataset}{Schlagenhauf, T.; Landwehr, M.; Fleischer, J. (2021): Industrial Machine Tool Element Surface Defect Dataset. \textit{Karlsruhe Institute of Technology (KIT)}. DOI: 10.5445/IR/1000129520}{https://publikationen.bibliothek.kit.edu/1000129520}{Mechanical component}{Ball screw drive}{Prognosis}{image}

\myRow{33}{Mendeley - Battery Degradation Dataset (Fixed Current Profiles \& Arbitrary Uses Profiles)}{Lu, Jiahuan; Xiong, Rui; Tian, Jinpeng; Wang, Chenxu; Hsu, Chia-Wei; Tsou, Nien-Ti; Sun, Fengchun; Li, Ju (2021), “Battery Degradation Dataset (Fixed Current Profiles \& Arbitrary Uses Profiles)”, \textit{Mendeley Data}, V2, doi: 10.17632/kw34hhw7xg.2}{https://data.mendeley.com/datasets/kw34hhw7xg/2}{Electrical component}{Battery}{Prognosis}{current, temperature, voltage}

\myRow{34}{Mendeley - Brushless DC motor}{Mazzoleni, Mirko; Scandella, Matteo; Previdi, Fabio; Pispola, Giulio (2019). Data for: First endurance activity of a Brushless DC motor for aerospace applications - REPRISE project, Mendeley Data, V2, doi: 10.17632/m58bdhy2df.2}{https://data.mendeley.com/datasets/m58bdhy2df/2}{Drive technology}{Brushless DC Motor}{Prognosis}{current, position, temperature}

\myRow{35}{Mendeley - Data for: Accelerated Cycle Life Testing and Capacity Degradation Modeling of LiCoO2-graphite Cells
}{Diao, Weiping (2019), “Data for: Accelerated Cycle Life Testing and Capacity Degradation Modeling of LiCoO2-graphite Cells”, \textit{Mendeley Data}, V1,DOI: 10.17632/c35zbmn7j8.1}{https://data.mendeley.com/datasets/c35zbmn7j8/1}{Electrical component}{Battery}{Prognosis}{capacity}

\myRow{36}{Mendeley - Diesel Engine Faults}{Denys Pestana-Viana - Federal Center of Technological Education Celso Suckow da Fonseca (CEFET-RJ), Rio de Janeiro, Brazil}{https://data.mendeley.com/datasets/k22zxz29kr/1}{Drive technology}{Diesel Engine}{Diagnosis}{pressure, temperature, vibration}

\myRow{37}{Mendeley - HUST Bearing}{Hong, Hoang Si; Thuan, Nguyen (2023), “HUST bearing: a practical dataset for ball bearing fault diagnosis”, Mendeley Data, V3, doi: 10.17632/cbv7jyx4p9.3}{https://data.mendeley.com/datasets/cbv7jyx4p9/3}{Mechanical component}{Bearing}{Diagnosis}{vibration}

\myRow{38}{Mendeley - Long-term Dynamic Durability Test Dataset for Single Proton Exchange Membrane Fuel Cell}{Zuo, Jian (2024), “Long-term Dynamic Durability Test Dataset for Single Proton Exchange Membrane Fuel Cell”, \textit{Mendeley Data}, V1, doi: 10.17632/w65jjt8v5w.1}{https://data.mendeley.com/preview/w65jjt8v5w?a=5b37947d-ea28-48cd-a5f2-d8c61c8ed8b1}{Electrical component}{Fuel Cell}{Prognosis}{unknown}

\myRow{39}{Mendeley - NMC cell 2600 mAh cyclic aging data}{Burzyński, Damian; Kasprzyk, Leszek (2021), “NMC cell 2600 mAh  cyclic aging data”, \textit{Mendeley Data}, V1, doi: 10.17632/k6v83s2xdm.1}{https://data.mendeley.com/datasets/k6v83s2xdm/1}{Electrical component}{Battery}{Prognosis}{current, temperature}

\myRow{40}{Mendeley - Run-to-Failure Vibration Dataset of Self-Aligning Double-Row Ball Bearings}{Gabrielli, Alberto; Battarra, Mattia; Mucchi, Emiliano; Dalpiaz, Giorgio (2024). Physics-based prognostics of rolling-element bearings: The equivalent damaged volume algorithm, \textit{Mechanical Systems and Signal Processing}, Volume 215, 2024, 111435, doi: 10.1016/j.ymssp.2024.111435.}{https://data.mendeley.com/datasets/htk59pp5wx/1}{Mechanical component}{Bearing}{Prognosis}{vibration}

\myRow{41}{MFPT - Condition Based Maintenance Fault}{Data Assembled and Prepared on behalf of MFPT by Dr Eric Bechhoefer, Chief Engineer, NRG Systems}{https://www.kaggle.com/datasets/emperorpein/mfpt-fault-datasets}{Mechanical component}{Bearing}{Diagnosis}{vibration}

\myRow{42}{NASA - Accelerated Battery Life Testing}{Fricke, K., Nascimento, R., Corbetta, M., Kulkarni, C., Viana, F. (2023). Prognosis of Li-ion Batteries Under Large Load Variations Using Hybrid Physics-Informed Neural Networks. \textit{Proceedings of the Annual Conference of the PHM Society}, 15(1). doi: 10.36001/phmconf.2023.v15i1.3463}{https://www.nasa.gov/intelligent-systems-division/discovery-and-systems-health/pcoe/pcoe-data-set-repository/}{Electrical component}{Battery}{Prognosis}{current, temperature, voltage}

\myRow{43}{NASA - Bearing Data Set}{J. Lee, H. Qiu, G. Yu, J. Lin, and Rexnord Technical Services. IMS, University of Cincinnati. ”Bearing Data Set”, NASA Ames Prognostics Data Repository, NASA Ames Research Center, Moffett Field, CA}{https://www.nasa.gov/intelligent-systems-division/discovery-and-systems-health/pcoe/pcoe-data-set-repository/}{Mechanical component}{Bearing}{Prognosis}{vibration}

\myRow{44}{NASA - Capacitor Electrical Stress}{J. Renwick, C. Kulkarni, and J Celaya ”Capacitor Electrical Stress Data Set”, NASA Ames Prognostics Data Repository, NASA Ames Research Center, Moffett Field, CA}{https://www.nasa.gov/intelligent-systems-division/discovery-and-systems-health/pcoe/pcoe-data-set-repository/}{Electrical component}{Electrolytic Capacitors}{Prognosis}{impedance, voltage}

\myRow{45}{NASA - Capacitor Electrical Stress 2}{J. Celaya, C. Kulkarni, G. Biswas, and K. Goebel ”Capacitor Electrical Stress Data Set - 2”, NASA Ames Prognostics Data Repository, NASA Ames Research Center, Moffett Field, CA}{https://www.nasa.gov/intelligent-systems-division/discovery-and-systems-health/pcoe/pcoe-data-set-repository/}{Electrical component}{Electrolytic Capacitor}{Prognosis}{capacity, resistance}

\myRow{46}{NASA - CFRP Composites Data Set}{Abhinav Saxena, Kai Goebel, Cecilia C. Larrosa, and Fu-Kuo Chang ”CFRP Composites Data Set”, NASA Ames Prognostics Data Repository, NASA Ames Research Center, Moffett Field, CA}{https://www.nasa.gov/intelligent-systems-division/discovery-and-systems-health/pcoe/pcoe-data-set-repository/}{Materials}{Carbon Fibre-Reinforced Polymer Composite}{Prognosis}{operating condition, strain, vibration, x-ray}

\myRow{47}{NASA - HIRF Battery}{C. Kulkarni, E. Hogge, C. Quach and K. Goebel ”HIRF Battery Data Set”, NASA Ames Prognostics Data Repository
(http://ti.arc.nasa.gov/project/prognosticdata-repository), NASA Ames Research Center, Moffett Field, CA}{https://www.nasa.gov/intelligent-systems-division/discovery-and-systems-health/pcoe/pcoe-data-set-repository/}{Electrical component}{Battery}{Prognosis}{current, temperature, voltage}

\myRow{48}{NASA - IGBT}{J. Celaya, Phil Wysocki, and K. Goebel ”IGBT Accelerated Aging Data Set”, NASA Ames Prognostics Data Repository, NASA Ames Research Center, Moffett Field, CA}{https://www.nasa.gov/intelligent-systems-division/discovery-and-systems-health/pcoe/pcoe-data-set-repository/}{Electrical component}{Transistor}{Prognosis}{current, temperature, voltage}

\myRow{49}{NASA - Li-ion Battery Aging Datasets}{B. Saha and K. Goebel. ”Battery Data Set”, NASA Ames Prognostics Data Repository, NASA Ames Research Center, Moffett Field, CA}{https://www.nasa.gov/intelligent-systems-division/discovery-and-systems-health/pcoe/pcoe-data-set-repository/}{Electrical component}{Battery}{Prognosis}{current, temperature, voltage}

\myRow{50}{NASA - Milling Data set}{A. Agogino and K. Goebel. BEST lab, UC Berkeley. ”Milling Data Set ”, NASA Ames Prognostics Data Repository, NASA Ames Research Center, Moffett Field, CA}{https://www.nasa.gov/intelligent-systems-division/discovery-and-systems-health/pcoe/pcoe-data-set-repository/}{Manufacturing process}{Milling}{Prognosis}{acoustic emission, current, vibration}

\myRow{51}{NASA - MOSFET Thermal Overstress Aging}{J. R. Celaya, A. Saxena, S. Saha, and K. Goebel ”MOSFET Thermal Overstress Aging Data Set”, NASA Ames Prognostics Data Repository, NASA Ames Research Center, Moffett Field, CA}{https://www.nasa.gov/intelligent-systems-division/discovery-and-systems-health/pcoe/pcoe-data-set-repository/}{Electrical component}{Transistor}{Prognosis}{current, temperature, voltage}

\myRow{52}{NASA - Randomized Battery Usage Data Set}{B. Bole, C. Kulkarni, and M. Daigle ”Randomized Battery Usage Data Set”, NASA Ames Prognostics Data Repository, NASA Ames Research Center, Moffett Field, CA}{https://www.nasa.gov/intelligent-systems-division/discovery-and-systems-health/pcoe/pcoe-data-set-repository/}{Electrical component}{Battery}{Prognosis}{current, temperature, voltage}

\myRow{53}{NASA - Small Satellite Power Simulation}{C. Kulkarni and A. Guarneros ”Small Satellite Power Simulation Data Set”, NASA Ames Prognostics Data Repository, NASA Ames Research Center, Moffett Field, CA}{https://www.nasa.gov/intelligent-systems-division/discovery-and-systems-health/pcoe/pcoe-data-set-repository/}{Electrical component}{Battery}{Diagnosis}{current, temperature, voltage}

\myRow{54}{NASA - Turbofan engine degradation simulation data set}{A. Saxena and K. Goebel. ”Turbofan Engine Degradation Simulation Data Set”, NASA Ames Prognostics Data Repository, NASA Ames Research Center, Moffett Field, CA}{https://www.nasa.gov/intelligent-systems-division/discovery-and-systems-health/pcoe/pcoe-data-set-repository/}{Drive technology}{Aircraft Engine*}{Prognosis}{anonymized, operating condition}

\myRow{55}{NASA - Turbofan engine degradation simulation data set 2}{M. Chao, C.Kulkarni, K. Goebel and O. Fink (2021). ”Aircraft Engine Run-to-Failure Dataset under real flight conditions”, NASA Ames Prognostics Data Repository, NASA Ames Research Center, Moffett Field, CA}{https://www.nasa.gov/intelligent-systems-division/discovery-and-systems-health/pcoe/pcoe-data-set-repository/}{Drive technology}{Aircraft Engine*}{Prognosis}{anonymized, operating condition}

\myRow{56}{NIST - Robot Arm Position Accuracy}{National Institute of Standards and Technology NIST (USA), Engineering Laboratory, Intelligent Systems Division}{https://www.nist.gov/el/intelligent-systems-division-73500/degradation-measurement-robot-arm-position-accuracy}{Robotic}{Industrial Robot}{Diagnosis}{current, position, speed}

\myRow{57}{OpenEI - Evaluation of Building Fault}{Lin Guanjing and Robin Mitchel, Lawrence Berkeley National Laboratory, (2019)}{https://data.openei.org/submissions/910}{Building}{Buildings}{Diagnosis}{pressure, speed, temperature}

\myRow{58}{OpenEI - Gearbox Fault Diagnosis}{Yogesh Pandy, National Renewable Energy Laboratory}{https://data.openei.org/submissions/623}{Mechanical component}{Gear}{Fault detection}{vibration}

\myRow{59}{OSF - Second-life lithium-ion battery aging dataset based on grid storage cycling}{Khan, M. A., \& Onori, S. (2025). Second-life lithium-ion battery aging dataset based on grid storage cycling. https://doi.org/10.17605/OSF.IO/8JNR5}{https://osf.io/8jnr5/overview}{Electrical component}{Battery}{Prognosis}{current, impedance, voltage}

\myRow{60}{Oxford - Oxford Battery Degradation Dataset}{Birkl, C. (2017). Oxford Battery Degradation Dataset 1. \textit{University of Oxford}. DOI: 10.5287/bodleian:KO2kdmYGg

}{https://ora.ox.ac.uk/objects/uuid:03ba4b01-cfed-46d3-9b1a-7d4a7bdf6fac}{Electrical component}{Battery}{Prognosis}{current, temperature, voltage}

\myRow{61}{PHM Data Challenge 2008 - Turbofan}{NASA Prognostics Center of Excellence}{https://www.nasa.gov/intelligent-systems-division/discovery-and-systems-health/pcoe/pcoe-data-set-repository/}{Drive technology}{Aircraft Engine*}{Prognosis}{anonymized, operating condition}

\myRow{62}{PHM Data Challenge 2009 - Gearbox Fault Detection}{PHM Society, Gearbox fault detection data set, 2010}{https://c3.nasa.gov/dashlink/resources/997/}{Mechanical component}{Gear System (Gears, Bearings, Shafts)}{Diagnosis}{speed, vibration}

\myRow{63}{PHM Data Challenge 2010 -  CNC milling machine cutters}{X. Li, B.S. Lim, J.H. Zhou, S. Huang, S.J. Phua, K.C. Shaw, and M.J. Er; Singapore Institute of Manufacturing Technology, 71 Nanyang Drive, Singapore 638075 School of Electrical and Electronic Engineering, Nanyang Technological University, Nanyang Avenue, Singapore 639798}{https://www.phmsociety.org/competition/phm/10}{Manufacturing process}{CNC Milling Machine Cutters}{Prognosis}{acoustic emission, force, vibration}

\myRow{64}{PHM Data Challenge 2011 - Anemometer Fault Detection}{Creators Unknown}{https://www.phmsociety.org/competition/phm/11/problem}{Mechanical component}{Anemometer}{Fault detection}{direction, speed, temperature}

\myRow{65}{PHM Data Challenge 2013 - Unknown}{Creators Unknown}{https://www.phmsociety.org/events/conference/phm/13/challenge}{Unknown}{unknown}{Health assessment}{anonymized}

\myRow{66}{PHM Data Challenge 2014 - Unknown}{Creators Unknown}{https://www.phmsociety.org/events/conference/phm/14/data-challenge}{Unknown}{unknown}{Health assessment}{anonymized}

\myRow{67}{PHM Data Challenge 2015 - Plant Fault Detection}{Creators Unknown}{https://www.phmsociety.org/events/conference/phm/15/data-challenge}{Production system}{Industrial Plant Monitoring}{Diagnosis}{anonymized}

\myRow{68}{PHM Data Challenge 2016 - Semiconductor CMP}{Crystec Technology Trading GmbH}{https://www.phmsociety.org/events/conference/phm/16/data-challenge}{Manufacturing process}{Chemical mechanical planarization system}{Prognosis}{flow rate, operating condition, pressure, speed}

\myRow{69}{PHM Data Challenge 2017 - Bogie Vehicle}{Creators Unknown}{https://www.phmsociety.org/events/conference/phm/17/data-challenge}{Mechanical component}{Bogie}{Diagnosis}{vibration}

\myRow{70}{PHM Data Challenge 2018 - Ion Mill in Wafer Manufacturing}{Kai Goebel (at this time at NASA)}{https://www.phmsociety.org/events/conference/phm/18/data-challenge}{Production system}{Ion Mill Etching Tool}{Prognosis}{current, pressure, speed}

\myRow{71}{PHM Data Challenge 2019 - Fatigue Cracks}{He J, Guan X, Peng T, Liu Y, Saxena A, Celaya J, Goebel K (2013). A multi-feature integration method for fatigue crack detection and crack length estimation in riveted lap joints using Lamb waves. \textit{Smart Materials and Structures}. 22(10):105007.}{https://c3.ndc.nasa.gov/dashlink/resources/1014/}{Materials}{Aluminum Structure with Dynamic Loading}{Prognosis}{vibration}

\myRow{72}{PHM Data Challenge 2020 Europe - Filtration System}{Eker, O. F., Camci, F., \& Jennions, I. K. (2016). Physics-based prognostic modelling of filter clogging phenomena. \textit{Mechanical Systems and Signal Processing}, 75, pp. 395-412.}{https://www.sciencedirect.com/science/article/pii/S0888327015005609}{Process technology}{Filtration}{Prognosis}{flow rate, pressure}

\myRow{73}{PHM Data Challenge 2021 - Turbofan 2}{M. Chao, C.Kulkarni, K. Goebel and O. Fink (2021). ”Aircraft Engine Run-to-Failure Dataset under real flight conditions”, NASA Ames Prognostics Data Repository, NASA Ames Research Center, Moffett Field, CA (additional validation data compared to data set \textit{NASA - Turbofan engine degradation simulation data set 2} )}{https://data.phmsociety.org/2021-phm-conference-data-challenge/}{Drive technology}{Aircraft Engine*}{Prognosis}{anonymized, operating condition}

\myRow{74}{PHM Data Challenge 2021 Europe: SCARA-robot}{Swiss Centre for Electronics and Microtechnology (CSEM)}{https://github.com/PHME-Datachallenge/Data-Challenge-2021}{Robotic}{Industrial Robot}{Diagnosis}{current, duration, heat slope, humidity, position, pressure, speed, temperature}

\myRow{75}{PHM Data Challenge 2022 - Rock Drills}{Epiroc Rock Drills AB,
\"{O}rebro, 702 25, Sweden and Link\"{o}ping University, Link\"{o}ping, 581 83, Sweden}{https://data.phmsociety.org/2022-phm-conference-data-challenge/}{Mechatronic system}{Rock Drills}{Diagnosis}{pressure}

\myRow{76}{PHM Data Challenge 2022 Europe - PCB Production line}{Danilo Giordano and Martino Trevisan, (2022). ”PHME Data Challenge”. European conference of the prognostics and health management society. / BITRON: Strada del Portone 95 - 10095 Grugliasco (TO), Italy, https://www.bitron.com/en}{https://github.com/PHME-Datachallenge/Data-Challenge-2022}{Production system}{PCB Production Line}{Diagnosis}{inspection data}

\myRow{77}{PHM Data Challenge 2023 - Gearbox}{PHM Society, Gearbox data set, 2023}{https://data.phmsociety.org/phm2023-conference-data-challenge/}{Mechanical component}{Gear}{Diagnosis}{speed, vibration}

\myRow{78}{PHM Data Challenge 2023 Asia Pacific - Experimental Propulsion System}{Japan Aerospace Exploration Agency (JAXA)}{https://data.phmsociety.org/phmap-2023-data-challenge/}{Drive technology}{Propulsion System*}{Health assessment}{pressure}

\myRow{79}{PHM Data Challenge 2024 - Helicopter Turbine Engines}{Prognostics and Health Management (PHM) Society}{https://data.phmsociety.org/phm2024-conference-data-challenge/}{Drive technology}{Helicopter Turbine Engine}{Fault detection}{power, speed, temperature, torque}

\myRow{80}{PHM Data Challenge 2025 - Aircraft Engines}{Creators Unknown, data generation:
NASA AGTF30 Simulation – MATLAB Executable Steady-State Solver and Linearization Tool for the AGTF30 Engine Simulation (MEXLIN-AGTF30), https://software.nasa.gov/software/LEW-20688-1. See also https://software.nasa.gov/software/LEW-19717-1.
MATLAB T-MATS Toolbox – https://ntrs.nasa.gov/api/citations/20180002976/downloads/20180002976.pdf}{https://data.phmsociety.org/phm-north-america-2025-conference-data-challenge/}{Drive technology}{Aircraft Engine*}{Prognosis}{operation conditions, pressure}

\myRow{81}{PHM Data Challenge 2025 Asia Pacific - Predicting Cutter Flank Wear}{PHM Asia Pacific}{https://phmap25.org/datachallenge.html\#data-sharing}{Manufacturing process}{Turning machine cutter flank wear}{Prognosis}{acoustic emission, vibration}

\myRow{82}{PHM Data Challenge 2026 Europe - Subway ticket validation door}{Soualhi, M. (2026): PHME Data Challenge. \textit{European conference of the prognostics and health management society}}{https://phm-europe.org/data-challenge}{Mechatronic system}{Validation door}{Prognosis}{current, position, voltage}

\myRow{83}{PHM IEEE Data Challenge 2012 - FEMTO Bearing Data Set}{FEMTO-ST Institute, Besançon, France}{https://github.com/wkzs111/phm-ieee-2012-data-challenge-dataset}{Mechanical component}{Bearing}{Prognosis}{temperature, vibration}

\myRow{84}{PHM IEEE Data Challenge 2014 - Fuel Cell}{FCLAB Research Federation (FR CNRS 3539, France)}{https://repository.lboro.ac.uk/articles/dataset/IEEE_2014_Data_Challenge_Data/3518141}{Electrical component}{Fuel Cell}{Prognosis}{current, current density, flow rate, hygrometry, pressure, temperature, voltage}

\myRow{85}{PHM IEEE Data Challenge 2023 - Planetary Gearbox}{M\"{a}lardalen University (MDU) / M\"{a}lardalen Industrial Technology Center (MITC)}{http://www.ieeereliability.com/PHM2023/assets/data-challenge.pdf}{Mechanical component}{Planetary Gear}{Diagnosis}{vibration}

\myRow{86}{PHM Society - Electromechanical Ball Screw Drive}{GE Research and University of Tennessee Knoxville in part by the Advanced Research Projects Agency-Energy (ARPA-E), U.S. Department of Energy, under Award Number DEAR0001290}{https://data.phmsociety.org/servomotor_dataset/}{Mechatronic system}{Ball Screw Drive* (Electromechanical)}{Diagnosis}{current, position, speed, torque}

\myRow{87}{ResearchGate - Driveline Unbalanced Shaft}{Giacomo Barbieri and David Sanchez-Londono and Laura Cattaneo and Luca Fumagalli and David Romero}{https://www.researchgate.net/publication/341220576_Dataset_A_Case_Study_for_Problem-based_Learning_Education_in_Fault_Diagnosis_Assessment}{Mechanical component}{Unbalanced shaft (driveline)}{Diagnosis}{vibration}

\myRow{88}{SDOL - Diagnostics 101 bearing data}{Kim, S., An, D., Choi, J-H. (2020). Diagnostics 101: A Tutorial for Fault Diagnostics of Rolling Element Bearing Using Envelope Analysis in MATLAB. \textit{Applied Sciences}. 10(20):7302. doi:10.3390/app10207302}{https://www.kau-sdol.com/bearing}{Mechanical component}{Bearing}{Diagnosis}{vibration}

\myRow{89}{SDOL - HS Gear}{Creators Unknown}{https://www.kau-sdol.com/kaug}{Mechanical component}{Bearing}{Fault detection}{vibration}

\myRow{90}{SDOL - KAU Gear}{Creators Unknown}{https://www.kau-sdol.com/kaug}{Mechanical component}{Bearing}{Diagnosis}{encoder}

\myRow{91}{Toyota Research Institute - Battery Cycle Life}{Severson, K.A., Attia, P.M., Jin, N. et al. Data-driven prediction of battery cycle life before capacity degradation. Nat Energy 4, 383–391 (2019). https://doi.org/10.1038/s41560-0190356-8}{https://data.matr.io/1/projects/5c48dd2bc625d700019f3204}{Electrical component}{Battery}{Prognosis}{current, temperature, voltage}

\myRow{92}{UBFC - AMPERE Detection and diagnostics of rotor and stator faults in rotating machines}{Moncef Soualhi (Franche-Comt\'{e} Electronique M\'{e}canique Thermique et Optique - Sciences et Technologies (UMR 6174) (Universit\'{e} de Franche-Comt\'{e})), Abdenour Soualhi (Laboratoire d’Analyse des Signaux et des Processus Industriels), Thi-Phuong Khanh Nguyen, Kamal Medjaher (both Laboratoire G\'{e}nie de production), Guy Clerc, Hubert Razik (both Laboratoire Amp\`{e}re), doi: 10.25666/DATAUBFC-2023-03-06-03}{https://search-data.ubfc.fr/FR-13002091000019-2023-03-06-03_AMPERE-Detection-and-diagnostics-of-rotor-and.html}{Drive technology}{Squirrel cage motor (Rotor, Stator, Bearing)}{Diagnosis}{current, speed, vibration, voltage}

\myRow{93}{UBFC - LASPI Detection and diagnostics of gearbox faults}{Moncef Soualhi (Franche-Comt\'{e} Electronique M\'{e}canique Thermique et Optique - Sciences et Technologies (UMR 6174) (Universit\'{e} de Franche-Comt\'{e})), Abdenour Soualhi (Laboratoire d’Analyse des Signaux et des Processus Industriels), Thi-Phuong Khanh Nguyen, Kamal Medjaher (both Laboratoire G\'{e}nie de production), Guy Clerc, Hubert Razik (both Laboratoire Amp\`{e}re), doi: 10.25666/DATAUBFC-2023-03-06 .}{https://search-data.ubfc.fr/FR-13002091000019-2023-03-06_LASPI-Detection-and-diagnostics-of-gearbox.html}{Mechanical component}{Bearing + Gear}{Diagnosis}{current, vibration, voltage}

\myRow{94}{UBFC - METALLICADOUR Detection and diagnostics of multi-axis robot faults}{Moncef Soualhi (Franche-Comt\'{e} Electronique M\'{e}canique Thermique et Optique - Sciences et Technologies (UMR 6174) (Universit\'{e} de Franche Comt\'{e})), Abdenour Soualhi (Laboratoire d’Analyse des Signaux et des Processus Industriels), Thi-Phuong Khanh Nguyen, Kamal Medjaher (both Laboratoire G\'{e}nie de production), Guy Clerc, Hubert Razik (both Laboratoire Amp\`{e}re).}{https://search-data.ubfc.fr/FR-13002091000019-2023-03-06-02_METALLICADOUR-Detection-and-diagnostics-of.html}{Robotic}{Multi-axes robot (cutting tool + axes drifts)}{Diagnosis}{current, force, position, torque, vibration}

\myRow{95}{UCI - AI4I 2020 Predictive Maintenance}{Matzka, S. (2020). “Explainable Artificial Intelligence for Predictive Maintenance Applications,” \textit{Proceedings of the Third International Conference on Artificial Intelligence for Industries (AI4I)}, pp. 69-74, doi: 10.1109/AI4I49448.2020.00023 .}{https://archive.ics.uci.edu/ml/datasets/AI4I+2020+Predictive+Maintenance+Dataset}{Production system}{Manufacturing Machine}{Diagnosis}{speed, temperature, torque, wear}

\myRow{96}{UCI - Condition Based Maintenance of Naval Propulsion Plants Data Set}{1: DIBRIS - University of Genoa 2: School of Marine Science and Technology, Newcastle University}{http://archive.ics.uci.edu/ml/datasets/Condition+Based+Maintenance+of+Naval+Propulsion+Plants}{Drive technology}{Naval Propulsion Plant*}{Prognosis}{flow rate, injection control, pressure, speed, temperature, torque}

\myRow{97}{UCI - Condition monitoring of hydraulic systems Data Set}{ZeMA - Center for Mechatronics and Automation Technology Saarbr\"{u}cken, Germany}{https://archive.ics.uci.edu/ml/datasets/Condition+monitoring+of+hydraulic+systems}{Mechatronic system}{Hydraulic System}{Diagnosis}{efficiency, flow rate, power, pressure, temperature, vibration}

\myRow{98}{UCI - Mechanical Analysis 1990}{F. Bergadano, A. Giordana, L. Saitta, (1990). University of Torino, Italy}{http://archive.ics.uci.edu/dataset/64/mechanical+analysis}{Mechatronic system}{Electromechanical Devices}{Diagnosis}{speed}

\myRow{99}{UCI - Robot Execution Failures Data Set}{Universidade Nova de Lisboa, Monte da Caparica, Portugal}{https://archive.ics.uci.edu/dataset/138/robot+execution+failures}{Robotic}{Industrial Robot}{Diagnosis}{force, torque}

\myRow{100}{UCI - Steel Plates Faults}{Semeion Research Center of Sciences of Communication - Rome, Italy}{https://archive.ics.uci.edu/ml/datasets/steel+plates+faults}{Materials}{Steel Plates Faults}{Diagnosis}{geometry, inspection data, operating condition}

\myRow{101}{Uni Lulea - Vibration Signals Wind Turbines}{Lulea University of Technology: Martin del Campo Barraza, Sergio (Department of Computer Science, Electrical and Space Engineering, Embedded Internet Systems Lab); Sandin, Fredrik (Department of Computer Science, Electrical and Space Engineering, Embedded Internet Systems Lab); Str\"{o}mbergsson, Daniel (Department of Engineering Sciences and Mathematics, Machine Elements)}{http://ltu.diva-portal.org/smash/record.jsf?pid=diva2\%3A1244889\&dswid=-6104}{Mechanical component}{Bearing (Wind Turbines)}{Fault detection}{vibration}

\myRow{102}{Uni Paderborn KAt - Bearing Damage}{University Paderborn Kat}{https://mb.uni-paderborn.de/kat/forschung/bearing-datacenter/data-sets-and-download\#c374354}{Mechanical component}{Bearing}{Diagnosis}{speed, temperature, torque}

\myRow{103}{Virkler - Fatigue Crack Propagation}{Virkler, D. A., Hillberry, B. M. and Goel, P. K. (1979). The Statistical Nature of Fatigue Crack Propagation. \textit{Journal of Engineering Materials and Technology}, vol. 101, 148–153.}{https://rdrr.io/github/SimoneHermann/hierRegSDE/man/Virkler.html}{Materials}{Aluminum Plate}{Prognosis}{crack length}

\myRow{104}{WASEDA - Fault Detection and Classification}{De Bruijn, B., Nguyen, T. A., Bucur, D., \& Tei, K. (2016). Benchmark datasets for fault detection and classification in sensor data. In A. Ahrens, O. Postolache, \& C. Benavente-Peces (Eds.), \textit{SENSORNETS 2016 - Proceedings of the 5th International Conference on Sensor Networks}, pp. 185-195. SciTePress. https://doi.org/10.5220/0005637901850195}{https://waseda.pure.elsevier.com/en/publications/benchmark-datasets-for-fault-detection-and-classification-in-sens}{Electrical component}{Sensors}{Diagnosis}{light, temperature}

\myRow{105}{Zenodo - Ball bearings subjected to time-varying load and speed conditions}{Aimiyekagbon, O. K. (2024). Run-to-failure data set of ball bearings subjected to time-varying load and speed conditions. Zenodo, doi: 10.5281/zenodo.10805043, University Paderborn}{https://zenodo.org/records/10868257}{Mechanical component}{Bearing}{Prognosis}{temperature, vibration}

\myRow{106}{Zenodo - Data-driven capacity estimation of commercial lithium-ion batteries from voltage relaxation}{Jiangong Zhu, Yixiu Wang, Yuan Huang, R. Bhushan Gopaluni, Yankai Cao, Michael Heere, Martin J. M\"{u}hlbauer, Liuda Mereacre, Haifeng Dai, Xinhua Liu, Anatoliy Senyshyn, Xuezhe Wei, Michael Knapp, \& Helmut Ehrenberg. (2022). Data-driven capacity estimation of commercial lithium-ion batteries from voltage relaxation [Data set]. Zenodo. https://doi.org/10.5281/zenodo.6405084}{https://zenodo.org/records/6405084}{Electrical component}{Battery}{Prognosis}{current, temperature, voltage}

\myRow{107}{Zenodo - Identifying degradation patterns of lithium ion batteries from impedance spectroscopy using machine learning}{Zhang, Y., Tang, Q., Zhang, Y., Wang, J., Stimming, U., \& Lee, A. A. (2020). Identifying degradation patterns of lithium ion batteries from impedance spectroscopy using machine learning [Data set]. \textit{Zenodo}. https://doi.org/10.5281/zenodo.3633835}{https://zenodo.org/records/3633835}{Electrical component}{Battery}{Prognosis}{current, temperature, voltage}

\myRow{108}{Zenodo - Monitoring data railway bridge Leuven}{Kristof Maes, \& Geert Lombaert. (2020). Monitoring data for railway bridge KW51 in Leuven, Belgium, before, during, and after retrofitting [Data set]. \textit{Zenodo}. https://doi.org/10.5281/zenodo.3745914}{https://zenodo.org/records/3745914}{Building}{Railway bridge}{Diagnosis}{acceleration, displacement, strain}

\myRow{109}{Zenodo - Predictive maintenance dataset}{Huawei Munich Research Center}{https://zenodo.org/record/3653909\#.YAmTTBYxkcQ}{Mechatronic system}{Elevator}{Prognosis}{humidity, vibration}

\myRow{110}{Zenodo - TBSI Sunwoda Battery Dataset}{Shengyu, T. (2024). TBSI Sunwoda Battery Dataset [Data set]. Zenodo. https://doi.org/10.5281/zenodo.10715209}{https://zenodo.org/records/10715209}{Electrical component}{Battery}{Prognosis}{current, voltage}

\end{xltabular}

\newpage
\section*{Appendix B: Classification of Publicly Available Data Sets}
\label{appendix_taxonomy}
\vspace{-10pt}
\begin{table*}[!h]
\centering
\caption{Classification of the \sumDataSets~publicly available data sets based on their domain, the respective application, and the addressed \ac{PHM} task (\acf{FD}, \acf{D}, \acf{HA}, and \acf{P}). An application marked with an asterisk (*) entails only simulated data sets and additionally the respective amount (n) per task (*\textsuperscript{n$\cdot$FD/D/HA/P}) if it entails partly simulated data sets.}
\label{table:taxonomyPubliclyDataSets}
\vspace{5pt}
\tabcolsep 8 pt
\begin{tabular}{c | c | c c c c || c}

\toprule
\multirow{1}{*}{\textbf{Domain}}     & \multirow{1}{*}{\textbf{Application}} & \multirow{1}{*}{\textbf{FD}} & \multirow{1}{*}{\textbf{D}} & \multirow{1}{*}{\textbf{HA}} & \multirow{1}{*}{\textbf{P}} & \textbf{Sum across domain} \\ 
\midrule[0.75pt]

Building                                 & Building                      &      & 1 &   &    & \multirow{2}{*}{2}   \\
                                         & Bridge                        &      & 1 &   &    &                      \\
\midrule
Drive technology                         & Aircraft engine*              &      &   &   & 5  & \multirow{5}{*}{12}  \\
                                         & Electric motor                & 1    & 1 &   & 1  &                      \\
                                         & Diesel engine                 &      & 1 &   &    &                      \\
                                         & Helicopter engine             & 1    &   &   &    &                      \\
                                         & Propulsion system*            &      &   & 1 & 1  &                      \\
\midrule
Electrical component                     & Battery                       &      & 1 &   & 15 & \multirow{6}{*}{25}  \\
                                         & Capacitor                     &      &   &   & 2  &                      \\
                                         & Circuit breaker               &      &   &   & 1  &                      \\
                                         & Fuel cell                     &      &   &   & 2  &                      \\
                                         & Sensor                        & 1    & 1 &   &    &                      \\
                                         & Transistor                    &      &   &   & 2  &                      \\
\midrule
Manufacturing process                    & Planarization system          &      &   &   & 1  & \multirow{5}{*}{8}   \\
                                         & Drilling                      &      & 1 &   &    &                      \\
                                         & Electrophoresis painting      &      &   &   & 1  &                      \\
                                         & Milling                       & 2    &   &   & 2  &                      \\
                                         & Turning                       &      &   &   & 1  &                      \\
\midrule
Material                                 & Aluminum plate                &      &   &   & 2  & \multirow{3}{*}{4}   \\
                                         & Polymer composite             &      &   &   & 1  &                      \\
                                         & Steel plate                   &      & 1 &   &    &                      \\
\midrule
Mechanical component                     & Anemometer                    & 1    &   &   &    & \multirow{6}{*}{26}  \\
                                         & Ball screw drive              & 1    &   &   & 1                         \\
                                         & Bearing*\textsuperscript{(1$\cdot$P)}
                                                                         & 2    & 8 &   & 7  &                      \\
                                         & Bogie                         &      & 1 &   &    &                      \\
                                         & Gear                          & 1    & 3 &   &    &                      \\
                                         & Shaft                         &      & 1 &   &    &                      \\
\midrule
Mechatronic system                       & Air compressor                &      & 1 &   &    & \multirow{8}{*}{8}   \\
                                         & Air pressure system           & 1    &   &   &    &                      \\
                                         & Electromechanical ball screw* &      & 1 &   &    &                      \\
                                         & Electromechanical device      &      & 1 &   &    &                      \\
                                         & Elevator                      &      &   &   & 1  &                      \\
                                         & Hydraulic system              &      & 1 &   &    &                      \\
                                         & Rock drill                    &      & 1 &   &    &                      \\
                                         & Validation door               &      &   &   & 1  &                      \\
\midrule
Process technology                       & Filtration                    &      &   &   & 5  & \multirow{2}{*}{7}   \\
                                         & Pump                          & 1    & 1 &   &    &                      \\
\midrule
Production system                        & Ion mill etching tool         &      &   &   & 1  & \multirow{4}{*}{8}   \\
                                         & Log data                      & 1    &   &   &    &                      \\
                                         & Production line               & 2    & 3 &   &    &                      \\
                                         & Shrink-wrapper                & 1    &   &   &    &                      \\
\midrule
Robotic                                  & Articulated robot             &      & 4 &   &    & \multirow{2}{*}{5}   \\
                                         & Linear robot                  &      & 1 &   &    &                      \\
\midrule
Unknown                                  & ---                           & 2    &   & 2 & 1  & 5                    \\

\midrule[0.75pt]
\multicolumn{2}{c|}{all} & \nrDataSetsFaultDetection & \nrDataSetsDiagnosis & \nrDataSetsHealthAssessment & \nrDataSetsPrognosis & \textbf{\sumDataSets} \\  \bottomrule

\end{tabular}


\end{table*}

\newpage
\section*{Appendix C: Signals within Publicly Available Data Sets}
\label{appendix_signals_data_sets}
\vspace{-10pt}
\begin{table*}[h]
\centering
\caption{Signals used in at least two of the given data sets. Sorted by the sum of signal occurrences across the tasks of \acf{FD}, \acf{D}, \acf{HA}, and \acf{P}.}
\label{table:taxonomySignalPubliclyDataSets}
\vspace{5pt}
\begin{tabular}{l | c c c c || c}
\toprule
\textbf{Signal}     &  \textbf{FD} & \textbf{D} & \textbf{HA} & \textbf{P} & \textbf{Sum across tasks} \\ 
\midrule[0.75pt]

Vibration            & 4  & 16 & 0  & 14 & 34 \\ \midrule
Current              & 2  & 9  & 0  & 22 & 33 \\ \midrule
Temperature          & 4  & 8  & 0  & 21 & 33 \\ \midrule
Voltage              & 2  & 4  & 0  & 19 & 25 \\ \midrule
Speed                & 4  & 11 & 0  & 5  & 20 \\ \midrule
Pressure             & 0  & 5  & 1  & 12 & 18 \\ \midrule
Anonymized\textsuperscript{(1)}
                     & 3  & 1  & 2  & 4  & 10 \\ \midrule
Flow Rate            & 0  & 1  & 0  & 9  & 10 \\ \midrule
Position             & 2  & 4  & 0  & 2  & 8  \\ \midrule
Torque               & 2  & 5  & 0  & 1  & 8  \\ \midrule
Operating Condition\textsuperscript{(2)}
                     & 0  & 1  & 0  & 6  & 7  \\ \midrule
Acoustic Emission    & 0  & 2  & 0  & 3  & 5  \\ \midrule
Force                & 0  & 3  & 0  & 1  & 4  \\ \midrule
Acceleration\textsuperscript{(3)}
                     & 1  & 2  & 0  & 0  & 3  \\ \midrule
Unknown\textsuperscript{(4)}
                     & 2  & 0  & 0  & 1  & 3  \\ \midrule
Capacity             & 0  & 0  & 0  & 2  & 2  \\ \midrule
Humidity             & 0  & 1  & 0  & 1  & 2  \\ \midrule
Image                & 1  & 0  & 0  & 1  & 2  \\ \midrule
Impedance            & 0  & 0  & 0  & 2  & 2  \\ \midrule
Inspection Data      & 0  & 2  & 0  & 0  & 2  \\ \midrule
Power                & 1  & 1  & 0  & 0  & 2  \\ \midrule
Strain               & 0  & 1  & 0  & 1  & 2  \\ \bottomrule

\multicolumn{6}{l}{\footnotesize(1): signal values changed/transformed, ~(2): varying operating conditions during the life} \\
\multicolumn{6}{l}{\footnotesize (3): in terms of movement, not vibration, (4): signal (or sensor) type unknown}

\end{tabular}
\end{table*}


\end{document}